# Physical discovery in representation learning via conditioning on prior knowledge: applications for ferroelectric domain dynamics

Yongtao Liu,[1] Bryan D Huey,[2] Maxim A. Ziatdinov,[1,3] and Sergei V. Kalinin[1,a]

[1] Center for Nanophase Materials Sciences, Oak Ridge National Laboratory, Oak Ridge, TN 37831, USA

[2] University of Connecticut, Materials Science and Engineering, Storrs, CT 06269, USA

[3] Computational Sciences and Engineering Division, Oak Ridge National Laboratory, Oak Ridge, Tennessee 37831, USA

Recent advances in electron, scanning probe, optical, and chemical imaging and spectroscopy yield bespoke data sets containing the information of structure and functionality of complex systems. In many cases, the resulting data sets are underpinned by low-dimensional simple representations encoding the factors of variability within the data. The representation learning methods seek to discover these factors of variability, ideally further connecting them with relevant physical mechanisms. However, generally the task of identifying the latent variables corresponding to actual physical mechanisms is extremely complex. Here, we explore an approach based on conditioning the data on the known (continuous) physical parameters, and systematically compare it with the previously introduced approach based on the invariant variational autoencoders. The conditional variational autoencoders (cVAE) approach does not rely on the existence of the invariant transforms, and hence allows for much greater flexibility and applicability. Interestingly, cVAE allows for limited extrapolation outside of the original domain of the conditional variable. However, this extrapolation is limited compared to the cases when true physical mechanisms are known, and the physical factor of variability can be disentangled in full. We further show that introducing the known conditioning results in the simplification of the latent distribution if the conditioning vector is correlated with the factor of variability in the data, thus allowing to separate relevant physical factors. We initially demonstrate this approach using 1D and 2D examples on a synthetic dataset and then extend it to the analysis of experimental data on ferroelectric domain dynamics visualized via Piezoresponse Force Microscopy.

[a] sergei2@ornl.gov

The tremendous success of physical sciences over the last two hundred years has been largely predicated on the search for and discovery of physical mechanisms, meaning simple laws and factors that can explain observations. The paradigmatic example of this, as eloquently summarized by Wigner in his oft-cited opinion,[1] is the discovery of Newton's laws. Similarly, while the correlative studies of celestial objects and their correlation with agriculture have been known since ancient Egypt and Sumer times, it was the discovery of the numerical laws by Kepler and Tycho Brage that laid the foundation for the models capable of predicting planetary motion. Such patterns are followed in most scientific fields still today, with the experimental observations used to derive correlative relationships that in turn underpin the emergence of physical models. These are often linked to symbolic regression, where simplicity and elegance of the mathematical law are considered as a strong indicator that the correct physical model has been identified. Overall, the greatest advantage of known symbolic or computational models (*e.g.* the lattice Hamiltonian in condensed matter physics[2,3] or force fields in molecular dynamics[4]) is their capability to extrapolate outside of the original measurement domain, predicting the effect of parameter changes, and generally allowing for interventional and counterfactual studies.[5] For example, Newton's laws allow predicting the trajectories of man-made objects, whereas modern calculation methods allow exploring properties and functionalities of not-yet realized molecules and materials.[6,7]

The rapid development of deep learning[8,9] methods over the last decade has provided a powerful new tool for physical research capable of building correlative relationships between multidimensional objects. While early applications have relied on purely correlative models, the developments over the last several years include the introduction of physical constraints and symmetries in the neural networks, making the interpolations consistent with prescribed physical models.[10] Similarly, the advancement in symbolic regression methods has allowed for the discovery of physical laws from observational data, first implemented in the framework of genetic algorithms[11] and subsequently extended towards deep learning symbolic regression methods,[10,12] Koopman operator based methods,[13-15] and Bayesian methods[16]. Much of this effort relied on the presence of robust physical descriptors, such as planetary coordinates in astronomy or atomic nuclei in electron microscopy studies.

However, in many cases accessible to observation are complex data sets representing static or dynamic fields, as exemplified by video data, atomic evolution movies in electron microscopy, and dynamic materials studies with Scanning Tunneling Microscopy (STM) and Scanning Probe Microscopy. In these cases, the presence of simple underlying physical mechanisms can also be postulated. For example, the contrast in STM is determined by the underlying atomic structure and associated spatial distribution of electronic densities, where the relationship between the two is defined by quantum mechanics. Similarly, the observed distribution of electromechanical activity on the surfaces of ferroelectric and ionic materials visualized by Piezoresponse Force Microscopy (PFM) is determined by local variations in materials functionalities. Effects of the image formation mechanisms are often non-negligible, sometimes masking or even inverting the measured parameters,[17] so also must be incorporated. Correspondingly, machine learning methods capable of physical discoveries from such data are of interest, including interpolating within and (à la Wigner) extrapolating outside of the original measurement domain.

Especially as experimental datasets continue to grow from a manageable handful to thousands of frames[18] and hundreds of millions of pixels or voxels,[19] recent advances in the generative statistical models such as simple and variational autoencoders (VAEs) [20-22] offer a pathway for addressing these problems. The general premise of the autoencoder is that the observational data set can be encoded via a small (compared to the intrinsic dimensionality of data) number of latent variables, where the relationship between the latent vector and the data object is defined by the encoder and decoder network. The multitude of available studies have illustrated that VAEs allow for the disentanglement of the latent representation, generally referring to the behavior where the variability along the selected latent variable corresponds to easily identifiable trends in data.[23-31] Naturally, this poses the challenge as to whether latent variables can be identified with specific physical mechanisms, or predefined or controlled.[32] Finally, of particular interest is whether generative models such as VAEs can be used to extrapolate outside the original distribution.

Here we explore the introduction of known physical mechanisms by conditional VAE, using the conditioning based on known (continuous) descriptors. We use the known or hypothesized physical factors as the condition and explore the complexity of resultant latent distributions as a measure of discovery. We further explore the potential of the conditional VAE approach to extrapolate outside of the original range of conditioning parameters. This approach is illustrated for model systems with known factors of variability, and further extended to experimental PFM data of ferroelectric domain dynamics.

The central idea of VAE is that high-dimensional real-space observations $x$ are produced by a relatively small number of latent variables $z$ representing the main factors of variation in the data. Hence, we are interested in finding the posterior $p(z|x)$ defined according to Bayes' theorem as

$$p(z|x) = \frac{p(x|z)p(z)}{p(x)} \quad (1)$$

where $p(x|z)$ is a likelihood function, $p(z)$ is a prior latent distribution, and $p(x) = \int p(z)p(x|z)dz$. Since the denominator is, in general, computationally intractable, the posterior $p(z|x)$ is approximated by a family of variational distributions $q(z|x)$ such that[20,21]

$$\log p(x) = \mathbb{E}_{q(z|x)}[\log p(x|z)] - D_{KL}(q(z|x)||p(z)) \quad (2)$$

where it is assumed that each datapoint is generated by a local latent variable. The first term is the expected conditional log-likelihood ('reconstruction error') and the second term $D_{KL}$ is the Kullback-Leibler divergence that forces the variational distribution $q(z|x)$ toward the latent prior $p(z)$. The latter is usually chosen to be the standard Gaussian distribution, in which case the KL term encourages the maximally 'simple' variational distributions. The $q(z|x)$ and $p(x|z)$ are commonly approximated by two neural networks called encoder $q_\phi(z|x)$ and decoder $p_\theta(x|z)$ where $\theta$ and $\phi$ are two sets of neural network weights optimized vi stochastic gradient descent. Computationally wise, the encoder takes data $x$ as input and outputs the parameters of the variational distribution. The decoder takes the latent variables sampled using the inferred distribution parameters and maps them back to the data space trying to reconstruct the original input.

The conditional VAE is an extension of the VAE that allows additionally conditioning the inference on the already known factors of variation.[33] In this case, the decoder's probability distribution is conditioned on the latent variable $z$ and some known parameters $c$, $p_\theta(z|x,c)$. Until now, most of the applications assumed $c$ to be a discrete variable, such as a digit label in MNIST. Here, we examine scenarios where $c$ represents a continuous parameter or set of parameters.

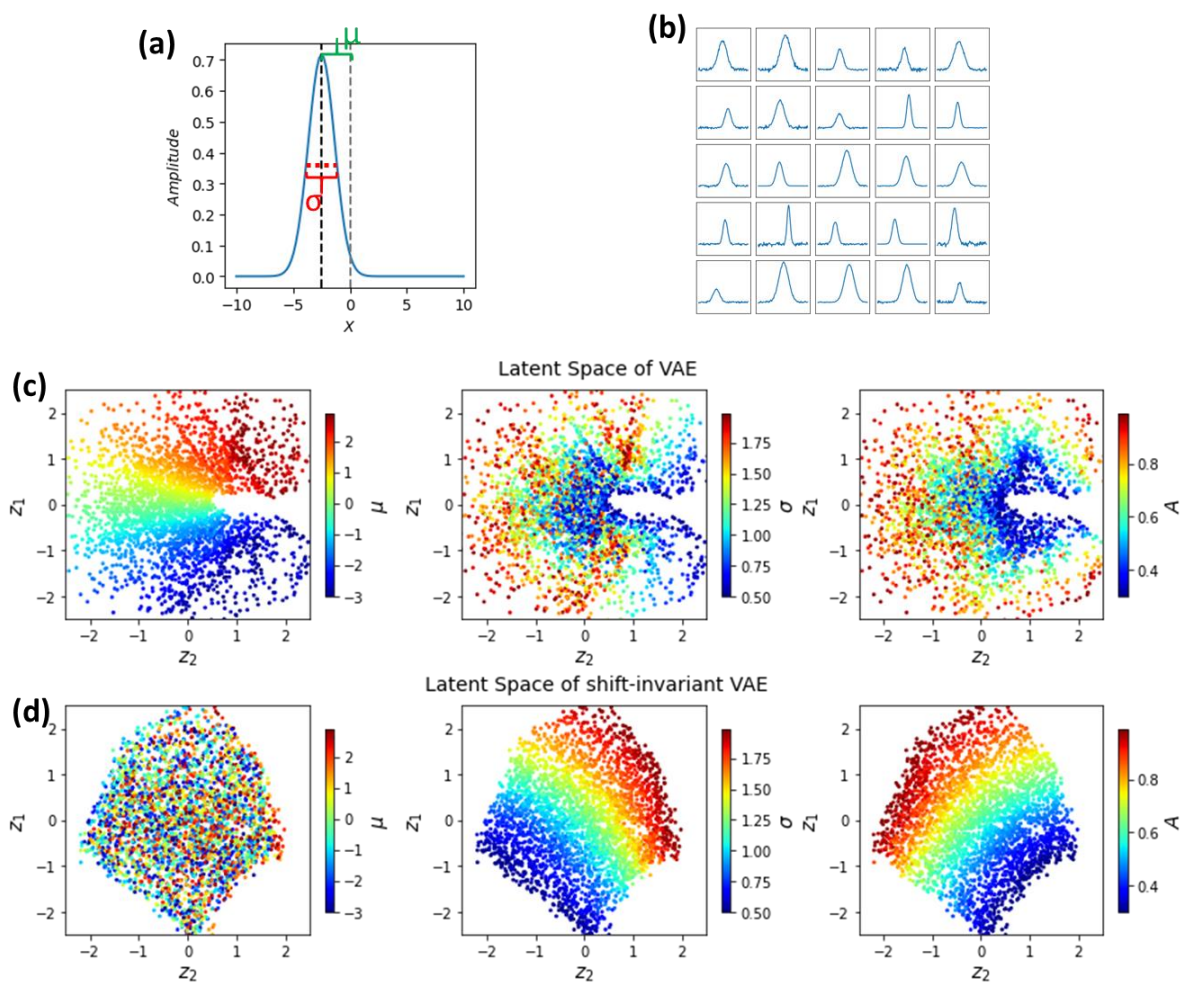

**Figure 1. Shift-invariant VAE analysis of 1D synthetic peak data. (a),** an example of 1D peak with labeled ground truth parameters, where $\mu$ is peak shift, $\sigma$ is peak width (full-width-half-max), and the peak maxima is the amplitude (A). **(b),** 36 randomly sampled examples out of 3000 in the synthetic peak data. **(c),** simple VAE of analysis of the peak data shown as latent space colored by ground truth parameters, where it is observed that all peak parameters are encoded into latent variables. **(d)**, Shift-invariant VAE analysis of the peak data shown as latent space colored by ground truth parameters, where there is no correlation between peak shift $\mu$ and latent variables because the peak shift is encoded into the shift variable as shown in **Figure S2d**.

To illustrate the principle of the cVAE operation and compare it with the previously introduced shift-invariant VAE, we developed a toy data set of Gaussian peaks with varying amplitudes, width, and positions. The toy data set includes $N = 3000$ Gaussian curves defined by function:

$$y = A \times e^{\frac{-(x-\mu)^2}{2\times\sigma^2}} + \delta_{noise} \qquad (3)$$

on $x \in [-10, 10]$, where $\mu$ is peak shift in the range of $[-3, 3]$, $A$ is the peak amplitudes in the range of $[0.3, 1]$, $\sigma$ is peak width in the range of $[0.5, 2]$, and $\delta_{noise}$ is the white noise with amplitude in the range of $[0, 0.03]$. Figures 1a-1b show example toy data. A Jupyter notebook is provided along with this manuscript allowing adjusting the toy data set parameters and subsequent analytics.

The simple VAE analysis of this dataset is shown in Figure 1 (c). Here, the latent distributions of the data are plotted with the color overlay corresponding to the ground truth labels. The latter are not available to the algorithm, and hence allow identification of the latent variables in terms of the data set parameters. Of course, the data has three factors of variability, and the latent space is two dimensional, thus we do not expect the full separation of the factors of variability. Still, examining the results clearly illustrates that the $z_1$ variable is largely associated with peak shift $\mu$, while the variability in the $z_2$ direction represents the joint effect of the amplitude $A$ and width $\sigma$.

Similar analysis using shift-invariant VAE encodes the data in terms of separating shift into a shift variable and the rest of information into standard latent variables. Examination of the data in Figure 1 (d) illustrates that now variability associated with the peak shift $\mu$ have disappeared, whereas variability in latent space represents the collective effect of amplitude and width. These findings are equivalent to our previous work and illustrate the capability of physically-defined invariances to disentangle them from real data.[31,34] However, these invariances are encoded in the coordinate transform in the invariant VAE (iVAE) framework, as described in depth in earlier works.[25-28,35-37]

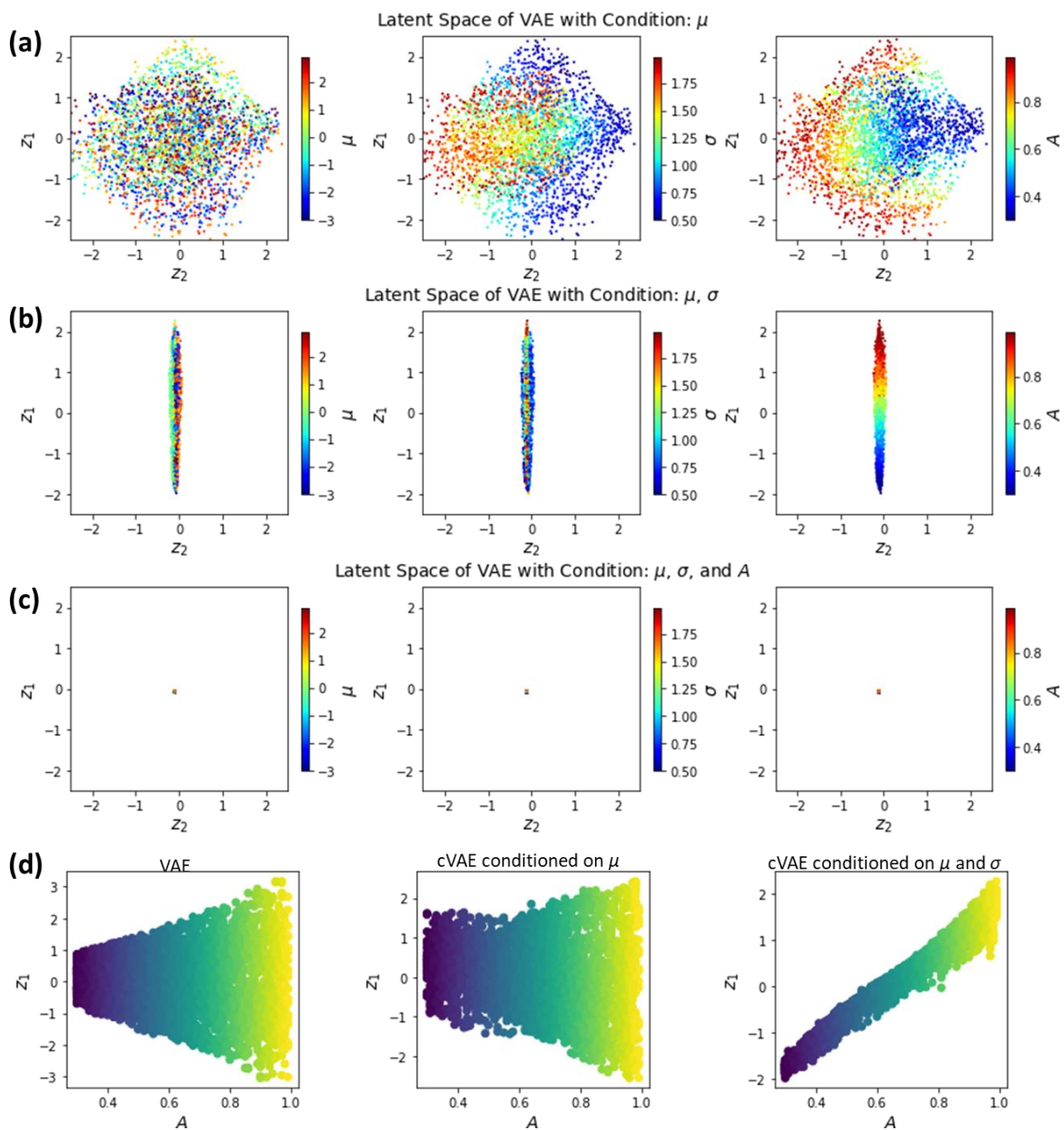

**Figure 2. cVAE analysis of 1D synthetic peak data.** (a) learned latent manifold of cVAE conditioned on the (known) peak shift $\mu$. In this case, there is no correlation between peak shift and latent variables, whereas both peak width $\sigma$ and peak amplitude $A$ are correlated with latent variable $z_2$. (b), learned latent manifold of cVAE conditioned on the peak shift $\mu$ and peak width $\sigma$. There is still no correlation between peak shift, peak width and latent variables, while peak amplitude $A$ is now correlated with latent variable $z_1$. (c), learned latent manifold of cVAE conditioned on the peak shift $\mu$, peak width $\sigma$, and peak amplitude $A$, where the latent manifold is completely collapsed because all peak parameters are added as conditions. (d) the correlation between $z_1$ and ground truth peak amplitude $A$ with different cVAE conditionings. It shows that conditioning simplifies analysis; that is, especially in the 3$^{rd}$ plot when conditioning on peak shift $\mu$ and width $\sigma$ (two of the three dataset variabilities), cVAE successfully encodes peak amplitude $A$ (the third variability) into $z_1$, evidenced by the linear correlation between $z_1$ and ground truth

peak amplitude $A$. More examples assessing latent variables versus ground truths are shown in Supplementary Information Figures S3-S5.

Next, we apply the cVAE approach to the same data set. In this case, the VAE receives the data set of the shape $N \times D$ and a conditional vector of the shape $N \times 1$ describing a known continuous parameter as an input. It is important to note that, unlike in iVAE models, there is no coordinate transform as a part of the model architecture, and the conditioning vector can represent any known salient feature or features. However, the use of the same physical parameters as for the toy model allows for physical comparisons.

The cVAE analysis conditioned on the peak shift $\mu$ is shown in Figure 2 (a). Examination of the ground truth labels illustrates that $z_2$ is still associated with peak amplitude $A$ and width $\sigma$, but not peak shift $\mu$. In this manner, the cVAE and shift-invariant VAE each lead to comparable outcomes. We then further explore conditioning on a pair of variables, namely $\mu$ and $\sigma$. In this case, the latent manifold is still 1D, but now the $z_1$ is clearly associated with the peak amplitude $A$. This is a demonstrable improvement over the shift-invariant VAE analysis, which did not allow for the separation of the two ground truths factors. Finally, the conditioning on all three variables results in the complete collapse of the representation, and the data manifold is now zero dimensional.

This simple 1D example demonstrates that conditioning on known factors of variability allows us to simplify the representations of the data and partially control the physical meaning of the remaining latent variables. As shown in Figure 2d, when conditioning on peak shift $\mu$ and width $\sigma$ (two of three variabilities of the dataset), cVAE clearly encoded peak amplitude $A$ (third variability of the dataset) into $z_1$. The associated changes in the dimensionality of the latent manifold thereby allow the number of intrinsic factors of variability in a data set to be explored.

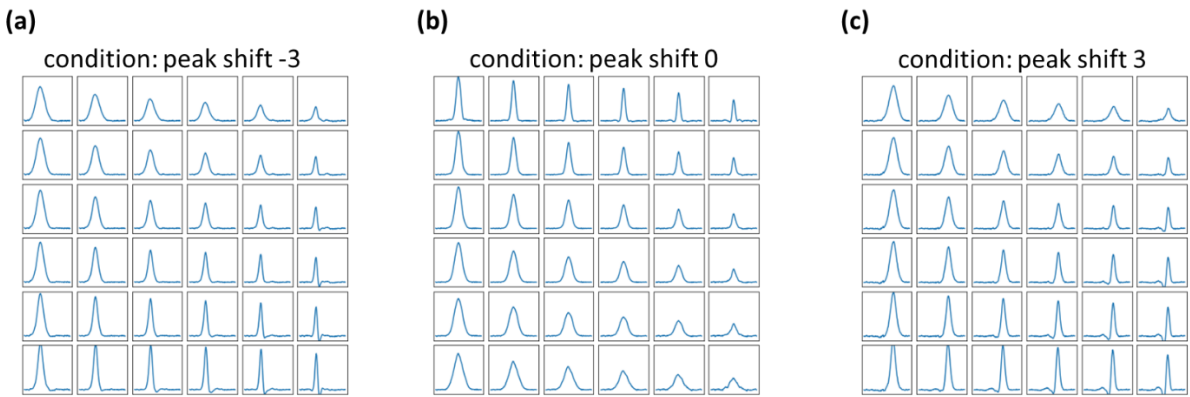

**Figure 3. Learned latent manifold of cVAE conditioned on a peak shift. (a), (b), (c),** latent manifold plots under various peak shift conditions. Note that each curve in the manifold shows a constant peak shift corresponding to the chosen condition (-3, 0, 3).

To continue the discussion, we also note that the trained cVAE model can be used to synthesize data with preserved latent traits allowing for interpolation and extrapolation along the conditioning parameter. This behavior is illustrated in Figure 3, which shows three latent manifolds produced by conditioning the trained cVAE's decoder on different peak shifts. For example, Figure 3a shows the manifold produced by conditioning the decoder on a peak shift of -3, resulting in a left-from-the-center shift of all peaks presented in Figure 3 (a); in the meantime, other peak parameters (*e.g.* peak width) still vary as expected. Similarly, manifolds in Figures 3 (b) and 3 (c) are conditioned on peak shifts of 0 and 3, respectively. Consequently, peaks in Figure 3 (b) are located at the center, and peaks in Figure 3 (c) are shifted to the right-hand side.

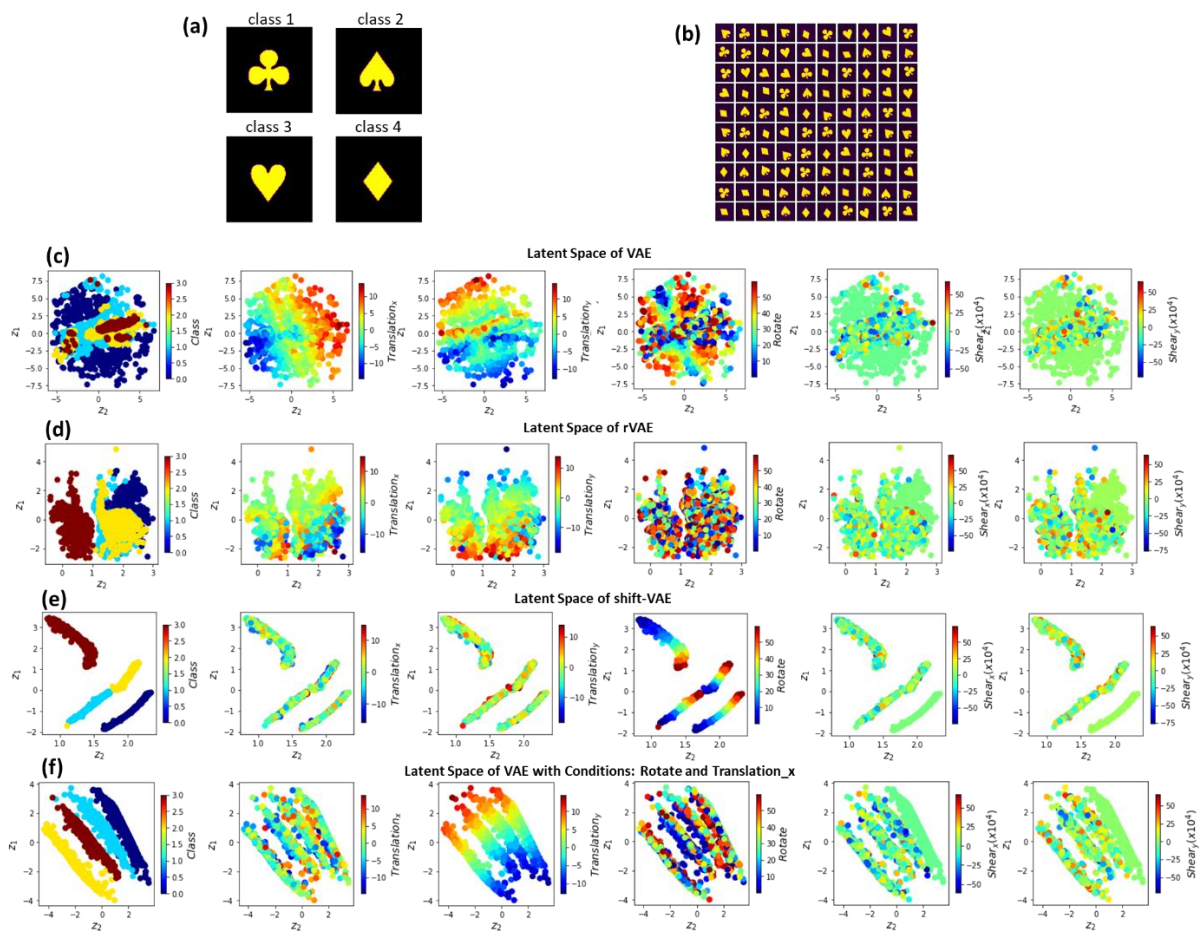

**Figure 4. Simple VAE, rVAE, shift-VAE, and cVAE analysis of 2D cards dataset. (a),** the source card images. **(b),** examples of the generated cards data with different shift, rotation, and shear. **(c),** latent space of simple VAE analysis, where clear correlations between (1) latent variables and class, (2) latent variables and *x*-translation, (3) latent variables and *y*-translation, (4) latent variables and rotation angle are observed. **(d)** latent space of rVAE analysis, where there is no obvious correlation between latent variables and rotation angle because rotation angle is encoded into rotation latent variable. **(e)** latent space of shift-VAE analysis, where there is no obvious correlation between latent variables and *x*-translation/*y*-translation because translations are encoded into the translation latent variable. **(f)** latent space of cVAE analysis conditioned on rotate and *x*-translation, where there is no obvious correlation between latent variables and rotation angle, as well as between latent variables and x-translation; in addition, (1) four classes are very obvious in latent space, (2) obvious correlation between latent variables and *y*-translation is seen; these indicate that cVAE simplifies the disentanglements of class and *y*-translation with known factor of variabilities, *i.e.*, rotation and *x*-translation. The plots of the discovered latent variables *vs* ground truth parameters for the VAE, rVAE, shift-VAE, and cVAE analyses in (**c**)-(**f**) are shown in Figures S6-S9, respectively. More analyses of cVAE with other conditions are shown in Figures S10-S12.

We further expand this approach to 2D objects. Here, we use the previously developed cards data set[30] that contains four classical card hands (as shown in Figure 4 (a)) augmented by rotations, translations, and shear. This data set allows for readily-identifiable discrete classes, as well as interesting degeneracies (*e.g.* rotated and deformed diamonds can be identical). The card dataset used here includes 4000 cards (1000 per each card suite) with various disorders including random rotations in the range of [0º, 60º], shifts in the range of [-4, 4] px, and shear in the range of [-0.002, 0.002]. Figure 4 (b) shows some example card images in the dataset.

We apply simple VAE, rVAE, shift-invariant VAE, and cVAE to this card dataset. The simple VAE analysis with the ground truth labels is illustrated in Figure 4 (c). Note that, due to close similarity between different cards after various rotations and deformations, the VAE fails to cluster the data set based on class variability and classes form connected manifolds. Rather, class-specific clusters form complex interpenetrating distributions in the latent space. That said, translation in *x* and *y* directions show clear alignment with chosen directions in the latent space. Similarly, rotation angle shows a complicated but perhaps distinguishable distribution. These complex behaviors emerge due to the competing tendencies of representation disentanglement when several physical factors of variability compete for representation by latent variables, giving rise to less than ideal latent distributions.

Shift-invariant VAE and rVAE allow to separate translation and rotation into specific latent variables, and the rest of the information is encoded into the standard latent variables. Figure 4 (d) shows the rVAE analysis results of the cards datasets. The rVAE reveals better performance for clustering the cards images based on class variability when the card rotation is separated into a rotation variable, though there is still a significant interpenetration between some classes. Then, $z_1$ is associated with both *x*-translation and *y*-translation. As expected, the rotation is not associate with $z_1$ and $z_2$ because it is encoded into the rotation variable. Figure 4 (e) shows even better performance on clustering the card images into four classes when implementing shift-VAE. As expected, there is no correlation between translations and standard latent variables as the translations are encoded into the translation variables. In this case, the rotation is associated with the standard latent variables $z_1$ and $z_2$.

Figure 4 (f) shows results from the cVAE analysis, where the data set was conditioned on both the rotation and *x*-translation. In this case, the latent space distribution clearly shows four unique clusters corresponding to the individual cards, with the variation between different classes associated with a selected direction in the latent space. The label distributions corresponding to the rotations and *x*-translation, *i.e.* to the variables on which conditioning has been performed, are featureless. At the same time, the translation in the *y*-direction, *i.e.* the only remaining factor of variability, becomes clearly associated with another direction in the latent space. Additional examples of cVAE analyses can be found in Figures S10-S12 in the supplementary materials, while the provided Jupyter notebook also allows more analyses to be explored. This illustrates that cVAE analysis with known physical parameters in the training session (*e.g.* rotation and *x*-translation) enables improving the disentanglement of latent representation of unknown physical parameters (*e.g. y*-translation).

It is also important to note that the behaviors of the latent representations of the data provide insight into physically relevant factors of variation within the data (*i.e.* classes, rotations, and translations). For invariant VAEs (rVAE and shift-invariant VAE), the introduction of invariances leads to the simplification of the latent distributions, which become controlled by the remaining (discrete or continuous) factors of variability. For the cVAE, introducing the known conditioning enables simplification of the latent distribution if the conditioning vector is correlated with the factor of variability in the data, thus allowing to separate relevant physical factors and experimental artefacts (*e.g.* scan distortions in experimental SPM data).

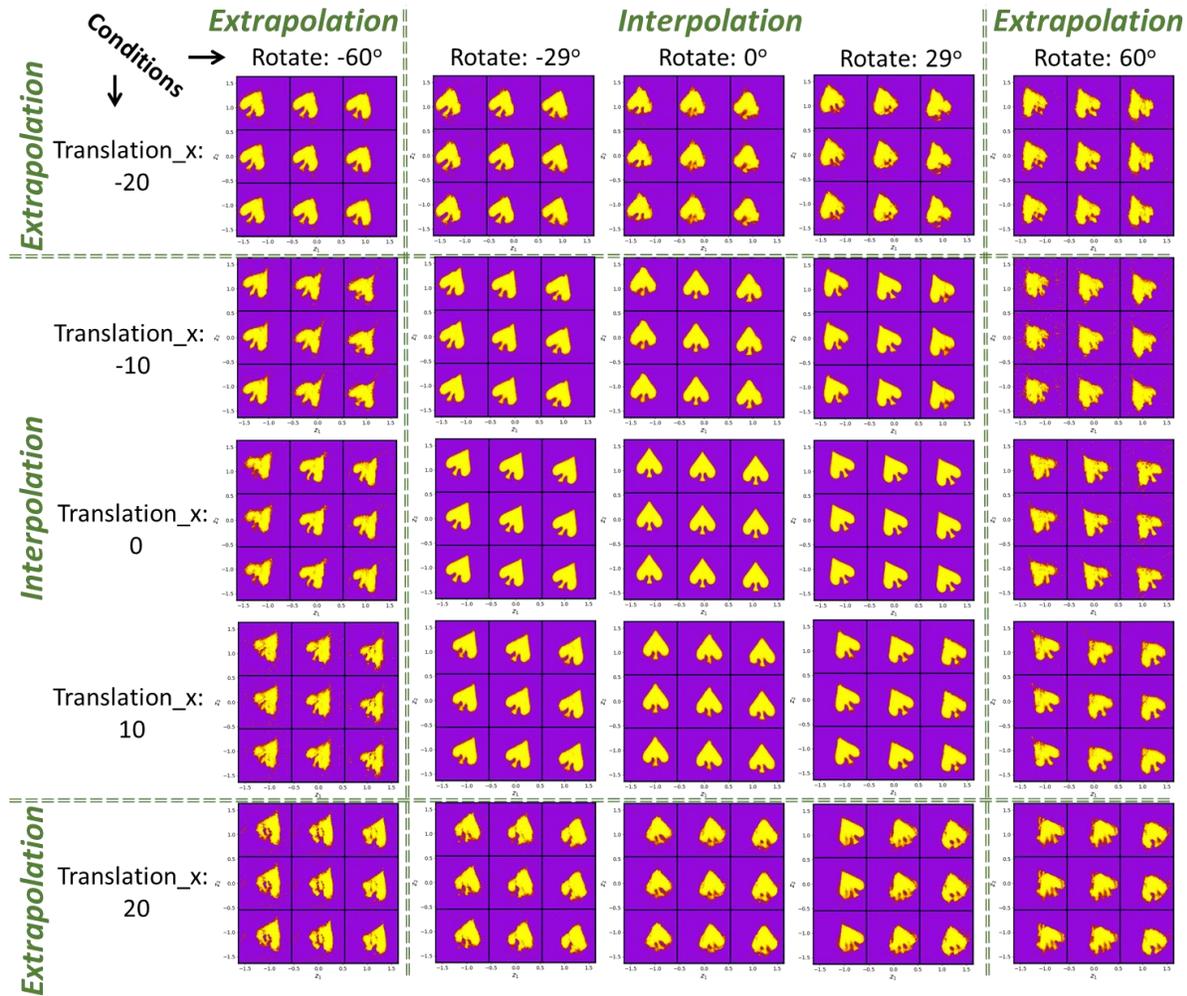

**Figure 5.** Learned latent manifolds of cVAE analysis of 2D cards data with rotation angle and *x*-translation as conditions. Each latent manifold shows constant *x*-translation and rotation corresponding to the defined conditions.

Similar to the 1D example, we further explore the potential of the cVAE to not just interpolate but also to extrapolate along the conditioning variables. Figure 5 demonstrates card images generated by a trained cVAE with rotation and *x*-translation as conditions. The training data includes 4000 cards with random rotations in the range of [0º, 60º] and shifts in the range of [-4, 4]. The extrapolation was performed by adding specified conditions to cVAE (*e.g. x*-translation=15, rotation=30). The extrapolated cards images shown in Figure 5 are consistent with the specified conditions. The card images in each sub-image shift from left-hand side to right-hand side of the field of view as the *x*-translation condition changes from -20 to 20. The tilt of the cards changes in a counterclockwise manner as the rotation condition changes from -60º to 60º. The chosen example also demonstrates the variability of decoded objects within the chosen regions of latent space. The extrapolation process is also available in the provided Jupyter notebooks allowing readers to explore it.

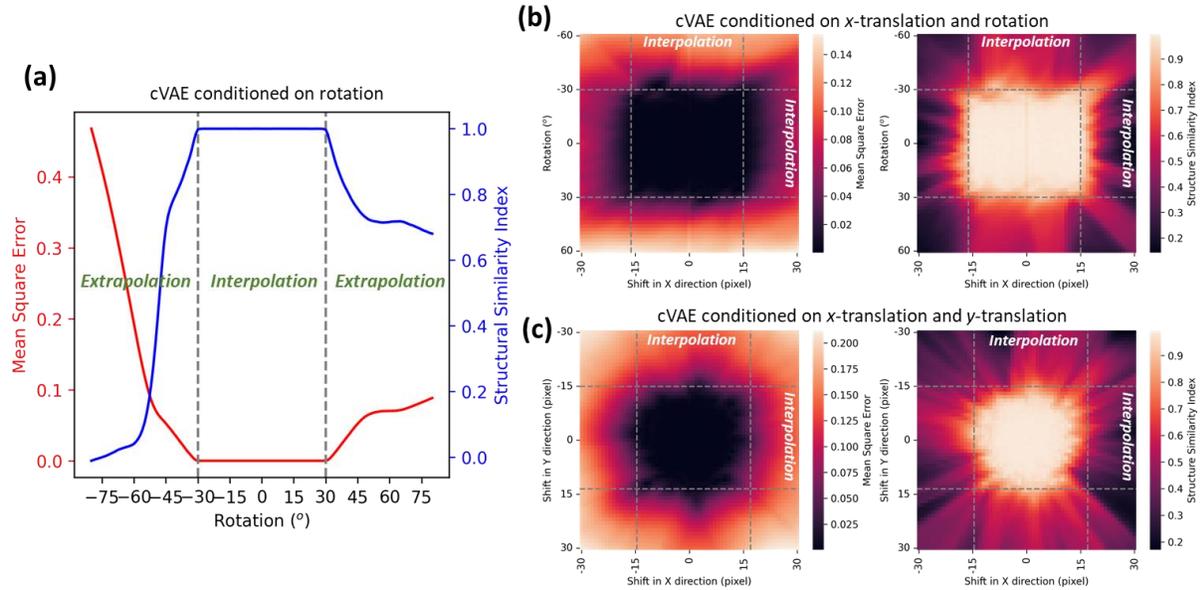

**Figure 6. Interpolation and extrapolation performance of cVAE** are shown as mean squared error and structure similarity index between the cVAE generated card data and the ground truth card data as a function of defined conditions. **(a)** performance of cVAE conditioned on rotation. **(b)** performance of cVAE conditioned on *x*-translation and rotation. **(c)** performance of cVAE conditioned on *x*-translation and *y*-translation.

In Figure 6, we summarized the interpolation and extrapolation performance by the mean squared error and structure similarity index between generated card images and the corresponding ground truth card image. We analyzed three models in Figure 6: (i) cVAE with rotation as a condition, (ii) cVAE with *x*-translation and rotation as conditions, (iii) cVAE with *x*-translation and *y*-translation as conditions. We explored the interpolation and extrapolation performance of all three models in Figure 6. Generally, interpolation performs extremely well, with almost perfect reconstruction within the training region. Figure 6 suggests that the cVAE allows for limited extrapolation on conditioning parameter. For the data set conditioned on the *x*-translation and rotation, the structure similarity index (SSID) for the images reconstructed well outside the original training region with the ground truth illustrating a clear matching pattern. Interestingly, the regions of good matching have a complex structure, where some directions in parameter space are associated with good interpolation whereas in other parts of parameter space the errors accumulate. These are unsurprising given the various local symmetries among the initial 4 cards. Similar behavior is observed for conditioning on *x*-translation and *y*-translation.

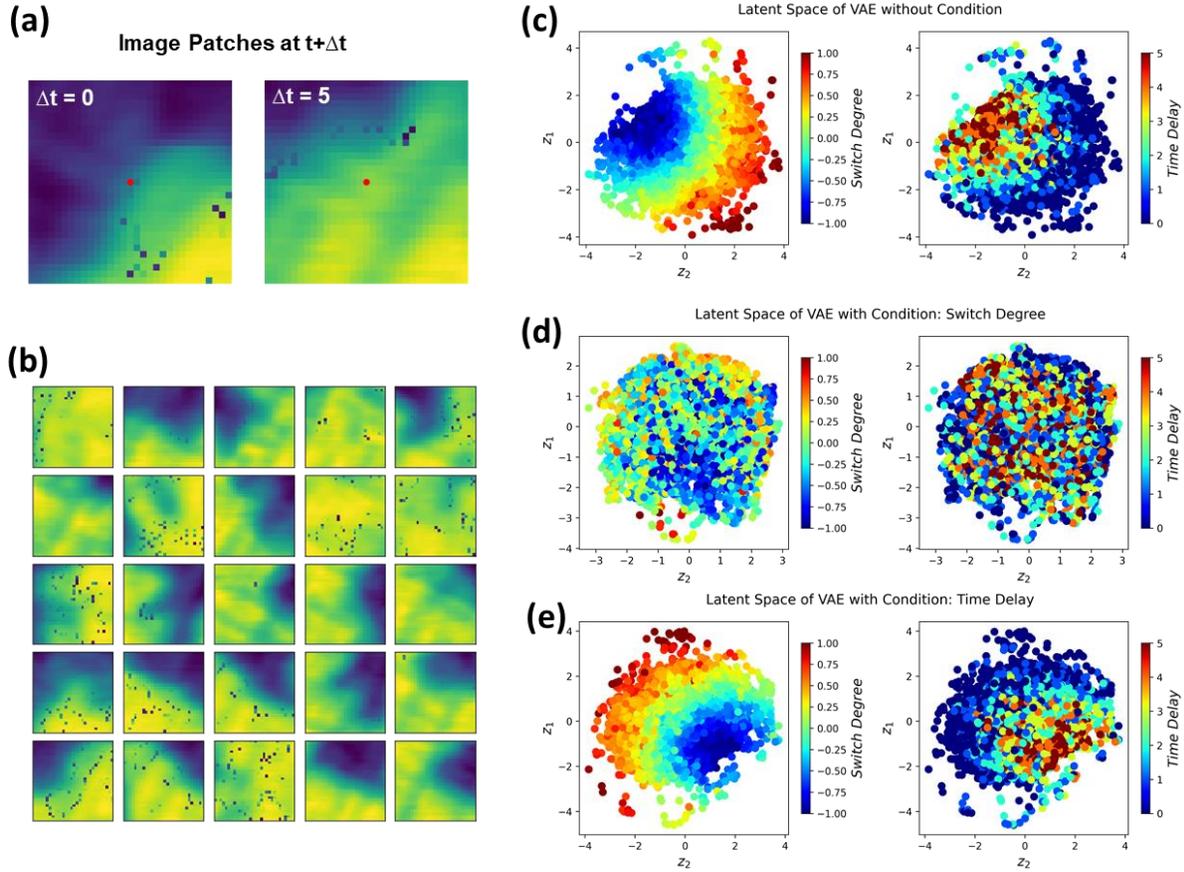

**Figure 7. Simple VAE and rVAE analysis with conditions of experimental PFM data. (a)** examples of PFM image patches (window size = 30) with different time delay. **(b)** examples of generated PFM image patches. **(c)-(e)** VAE analyses: **(c)** latent space of VAE analysis without conditions colored by switch degree and time delay, where a correlation between latent variables and switch degree, time delay is observed; **(d)** latent space of cVAE with switch degree as a condition, in this case, no correlation between ground truth parameters (switch degree and time delay) and latent variables is observed; **(e)** latent space of cVAE with time delay as a condition, in this case, the correlations between ground truth parameters (switch degree and time delay) and latent variables are modified but not disappear. It is seen that adding time delay as a condition is not functioning, this is probably because that there is conflict of information included in time delay and switch degree, as seen in (a), that is, larger time delay generally corresponds to larger switch degree. The latent variables vs ground truth parameters plots of these VAE analyses are shown in Figure S14.

With this thorough understanding of VAE based approaches, we extend this analysis to experimental data on ferroelectric domain switching. Previously, we have demonstrated the use of rVAE to explore ferroelectric domain switching pathway and domain wall dynamics. When applying rVAE to consecutive PFM images revealing the ferroelectric domain switching process, the polarization switching mechanism can be visualized in the latent space.[27] When applying rVAE to stacked ferroelectric and ferroelastic domain wall images (generated based on numerous continuously acquired PFM images during domain switching via electric field poling), it disentangles the factors affecting the ferroelectric domain wall dynamics. This includes how the distribution of ferroelastic domain walls affects the dynamics of ferroelectric domain walls, offering insight into the intrinsic mechanisms of ferroelectric polarization switching and hence approaches to engineer devices with more stable domains, domains that can switch faster, or for lower energy switching.[28] In particular, we probed the ferroelectric domain wall pinning mechanisms by translating the latent space to physical descriptors.[28] However, these analyses were enabled by the rotational invariances inherent to rVAE, but the physical interpretation of the latent variables was based exclusively on the analysis of the latent spaces. Here, we expand this analysis towards elucidation of the relevant latent mechanisms when the input data is conditioned on the *a priori* known physical descriptors.

As a model system, we explore the ferroelectric polarization switching dynamics in a 150 nm thick lead zirconate titanate (PZT) thin film grown on a $SrTiO_3$ (001) substrate by pulsed laser deposition (PLD), with a heteroepitaxial intermediate conducting oxide electrode (SRO).[38,39] We explore the domain switching dynamics as a function of time using PFM by applying a constant tip bias that just surpasses the coercive field. Consecutive PFM images (Figure S13) show the ferroelectric switching from the (001) to the (00-1) states. Consequently, domain switching can be excited and observed at the same time.[40,41] This PFM data was used in our earlier publication,[27] here we just reuse the PFM data to demonstrate the application of cVAE.

In the cVAE analysis, we introduced a time delay (*dt*). That means, the domain wall location is determined by a Canny filter at time *t*, the sub-image centered at the domain wall location is created at time *t* and *t+dt*. This leads to a comparison of domains at time *t* and *t+dt* in the sub-image datasets and hence the domain switching and wall dynamics are encoded as a dependent of time. Figure 7 (a) shows a comparison of domains at *dt=0* and *dt=5*, in the first image (*dt=0*) the domain wall is located at the center, in the second image (*dt=5*) the domain wall moves away from the center. Figure 7 (b) shows example sub-images used for cVAE analysis.

In cVAE analysis, we used switch degree and time delay as conditions. The switch degree represents the ratio of yellow (switched) domain and blue (unswitched) domain, and the time delay is explained above. Shown in Figure 7 (c) is a simple VAE analysis without conditions, where latent variables are colored by ground truth values. Just as with the 1D and 2D model systems explored above, the color gradient reveals that both switch degree and time delay are encoded into latent variables. Figures 7 (d) and 7 (e) then show cVAE analyses conditioned by switch degree and time delay, respectively. Figure 7 (d) confirms that the switch degree is featureless in the latent space when it is performed as a condition, indicating this prior physical knowledge effectively affects the cVAE analysis. However, the time delay is still visible as a correlation within the latent space when it is performed as a condition (Figure 7 (e)). Note that conditioning on switch degree not only leads to switch degree featureless in latent space but also leads to time delay featureless in latent space. This is possibly because of the intimate connection between switch degree and time delay, as shown in Figure 7 (a), when a time delay is added (*e.g. dt=5*), the switch degree also changes simultaneously.

We also explored the reconstruction of PFM image patches by cVAE with predefined parameters. Shown in Figure 8 are the analyses of two cVAE conditioned on time delay and switch degree, respectively. cVAE manifolds show the reconstruction of PFM image patches with different conditions. More reconstruction by cVAE is shown in Figure S15. Such cVAE manifolds as a function of defined conditions allow us to perform extrapolation into the future as well, as demonstrated in the final column in Figure 8. The analyses in Figure 7 and Figure 8 are extended to PFM domain wall images generated from raw PFM images via application of a Canny filter (results are shown in Figure S16-19).

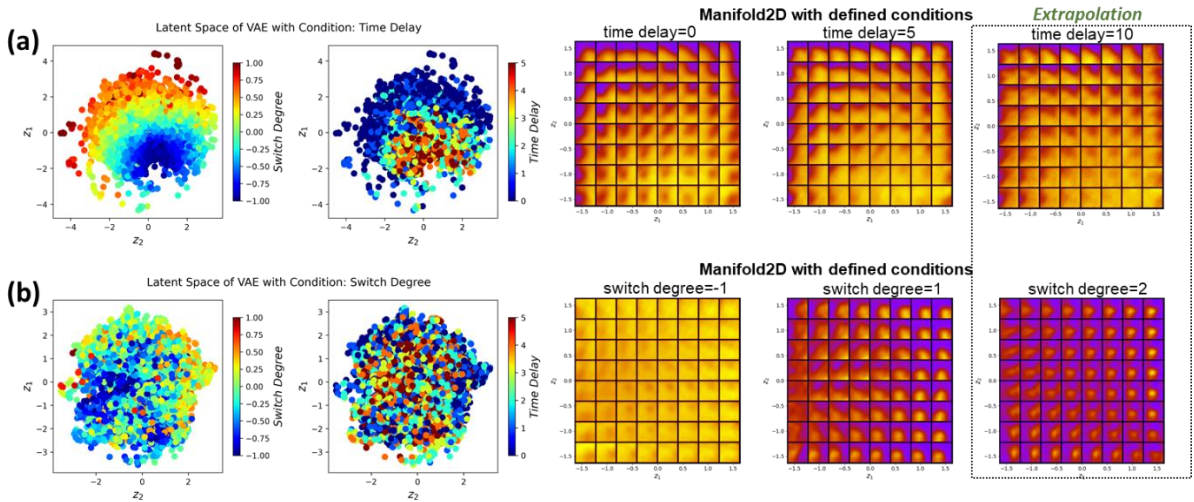

**Figure 8.** Latent manifolds with predefined conditions of cVAE analyses of experimental PFM data based on image patches with window size=20. **(a)** latent space distribution and latent manifold of VAE analysis with time-delay as a condition. **(b)** latent space distribution and latent manifold of VAE analysis with switch-degree as a condition.

In conclusion, we demonstrate the use of conditional variational autoencoders (cVAE) to explore physical information by conditioning prior known physical parameters, and we compare the cVAE with the previous invariant VAE (iVAE) approach. Given that cVAE does not rely on the specific invariant transform, it allows for much greater flexibility. We showed the reliability of this approach using modeled 1D spectrum and 2D image datasets, revealing that the conditioned parameters become featureless in the latent representation. Then, we extended this approach to experimental PFM data on ferroelectric domain switching and domain wall dynamics. We argue that the cVAE-based physics discovery can be performed in iterative and hypothesis testing modes, allowing for simple and low-dimensional latent distributions when the relevant physical factors are correctly identified.

## Methods

*PZT thin film*

The PZT film is grown on a SrTiO$_3$ (001) substrate by pulsed laser deposition (PLD), with an intermediate conducting oxide electrode (SRO). The PLD deposition is conducted at 650 °C with 100 mTorr oxygen partial pressure, then the samples are cooled to room temperature.

*Data analysis*

The detailed methodologies of cVAE and iVAE analysis are shown in Jupyter notebooks that are available from https://git.io/JD28J.

## Conflict of Interest

The authors declare no conflict of interest.

## Authors Contribution

S.V.K. conceived the project and M.Z. implemented the cVAE and iVAE codes via Pyro probabilistic programming language. Y.L. performed analyses. B.D.H performed PFM measurements of PZT samples. S.V.K. M.Z, and Y.L. wrote the manuscript. All authors contributed to discussions and the final manuscript.

## Acknowledgements

This effort (ML) is based upon work supported by the U.S. Department of Energy, Office of Science, Office of Basic Energy Sciences Energy Frontier Research Centers program under Award Number DE-SC0021118 (Y.L., S.V.K.), and the Oak Ridge National Laboratory's Center for Nanophase Materials Sciences (CNMS), a U.S. Department of Energy, Office of Science User Facility (M.Z.).

## Data and code availability:

The interactive Jupyter notebooks that reproduces this paper's results is available at https://git.io/JD28J.